\documentclass[11pt]{article} 
\usepackage[utf8]{inputenc}
\usepackage{natbib}
\usepackage[a4paper, total={6in, 8in}]{geometry} 

\usepackage{amsmath} 
\usepackage{amssymb}
\usepackage{graphicx}
\usepackage{caption}
\usepackage{subcaption}
\usepackage{wrapfig}

\title{Bayesian sense of time in biological and artificial brains}
\author{Zafeirios Fountas$^1$ \& Alexey Zakharov$^2$\\[2ex]
  \small To appear in: Time and Science (ed. R. Lestienne \& P. Harris)}
\date{
    \small $^1$Wellcome Centre for Human Neuroimaging, University College London,\\\vspace*{-0.5ex}
    \small 12 Queen Square, London WC1N 3AR, UK\\\vspace*{-0.5ex}
    \small z.fountas@ucl.ac.uk\\\vspace*{0.5ex} 
    \small $^2$Department of Computing, Imperial College London,\\\vspace*{-0.5ex}
    \small 180 Queen's Gate, London SW7 2RH, UK\\\vspace*{-0.5ex}
    \small az519@imperial.ac.uk\\[2ex]
    \small Draft compiled: \today
}

\begin{document}

\maketitle

\begin{abstract}
    Enquiries concerning the underlying mechanisms and the emergent properties of a biological brain have a long history of theoretical postulates and experimental findings. Today, the scientific community tends to converge to a single interpretation of the brain's cognitive underpinnings -- that it is a Bayesian inference machine. This contemporary view has naturally been a strong driving force in recent developments around computational and cognitive neurosciences. Of particular interest is the brain's ability to process the passage of time -- one of the fundamental dimensions of our experience. How can we explain empirical data on human time perception using the Bayesian brain hypothesis? Can we replicate human estimation biases using Bayesian models? What insights can the agent-based machine learning models provide for the study of this subject? In this chapter, we review some of the recent advancements in the field of time perception and discuss the role of Bayesian processing in the construction of temporal models. 
\end{abstract}

\section{Introduction}

More than 150 years ago, German polymath Hermann von Helmholtz has postulated that perception is realised by means of `unconscious inferences' which is learned over time rather than acquired at birth \citep{Helmholtz:1867}. Although this view was scrupulously criticised by the nativists of his time, it predicted scientific developments for over 100 years ahead and inspired the contemporary interpretations of the human brain's cognitive underpinnings. In particular, Helmholtz's view on perception is often cited as the earliest account of the brain being regarded as an inference machine and as one of the inspirations for the influential studies on generative models in the early 1990s \citep{Dayan1995TheHM, Friston2012TheHO}.

An illustrative example of this is a seminal paper by \cite{Dayan1995TheHM}, which begins with: ``Following Helmholtz, we view the human perceptual system as a statistical inference engine whose  function is to infer the probable causes of sensory input.'' This idea has gained traction in the second half of the 20th century, flourishing from the works on perception as hypothesis testing \citep{Gregory1968PerceptualIA, Kersten2004ObjectPA}, and establishing a strong case for the role of Bayesian probability theory in the fields of computational neuroscience and cognitive science \citep{Wolpert:1995, Knill:1996, Kording:2004}. The elegant amalgamation of generative models and Bayes' rule ultimately resulted in what is now known to be the `Bayesian brain' hypothesis \citep{TheBayesianBrain2004}. The hypothesis views the brain as a statistical machine equipped with a causal model of the environment, where perception is a consequence of Bayesian model inversion (probabilistic inference) given new sensory observations and the causal model is updated according to Bayes' rule \citep{Bubi2010PredictionCA, TheBayesianBrain2004}. 

Following these developments, a prominent theory of the brain's inner workings has emerged -- the free-energy principle \citep{Friston:2010:FEP}. At its core, the principle is simply a mathematical description underwriting the behaviour of any dynamical system separated by a Markov blanket from its environment \citep{Friston2019AFE}. In particular, it states that ``the  dynamics  of  persistent,  bounded  systems  may  be  framed  as  inferential  processes'', where ``states  internal  to  a  boundary  appear  to  infer  the  states  outside  of  it" \citep{Parr2019MarkovBI}. Attentive readers may notice the resemblance of this description with the Helmholtz-inspired research on generative models and the past interpretations of the brain as an inference machine. Indeed, one of the more ambitious goals of the free-energy principle is to provide a unifying theory for the brain's operations by explaining the fundamental tendencies of self-organising systems. Specifically, under the principle, the brain is a probabilistic (Bayesian) generative model, the objective of which is to minimise an information-theoretic quantity called the \textit{variational free energy} (also called negative marginal likelihood, or surprisal) \citep{Friston:2010:FEP}. Given the generality of the free-energy principle, it is consistent with a number of prominent research directions in the fields such as computational neuroscience (e.g. predictive coding) or machine learning (e.g. model-based reinforcement learning). With Bayesian mechanics at the core of the free-energy principle, the `Bayesian brain' hypothesis is similarly well accommodated within the boundaries of this grand theory.

As a Bayesian generative model, it is reasonable to assume that the brain must exhibit particular properties pertaining to its probabilistic nature -- as a corollary of Bayesian updating, the beliefs it embodies, and the predictions it makes based on them. Indeed, it has been observed that the brain can deal with uncertainty in a nearly optimal manner \citep{Clark:2015}, while Bayesian inference seems to have significant explanatory power for magnitude estimation by humans \citep{Petzschner:2015}. Magnitude estimation, in particular, is very telling as a conspicuous phenomenon that can be actively scrutinised through psychological experiments and computational models, in order to probe the hypothesis of Bayesian processing in the brain. 

Time, being one of such magnitudes susceptible to human estimation, has naturally become a subject of active research in the fields of computational neuroscience, cognitive science, and even agent-based machine learning. As one of the most basic dimensions in our quotidian existence, the means by which our brains represent, perceive and predict the passage of time remain an intriguing direction of work, particularly in light of the fairly recent theoretical and practical developments concerning the `Bayesian brain' hypothesis. In this chapter, we discuss how time perception can be formalised and studied from the Bayesian perspective using the latest literature on time perception models. Further, we dive deep into the state-of-the-art models proposed within the field of artificial intelligence, exploring a variety of ideas pertaining to how time can be represented in an artificial agent capable of intelligent behaviour.

\section{Interval timing}
Typically, the subjective perception of time is examined via interval timing tasks which, as the name suggests, assess the human (and other animals) ability to process the duration of (or between) events. Processing here refers to the ability to measure, estimate, produce or discriminate either isolated intervals or more complex temporal patterns \citep{Hardy:2016}. Although biological organisms have developed the ability to process time in multiple timescales, spanning over more than 10 orders of magnitude \citep{Buhusi:2005},  experiments of interval timing commonly focus on the range of milliseconds to minutes, due to the high tractability of this range in an experimental setting and its association with short-term brain dynamics and various aspects of behaviour, such as decision-making and motor control.

There is a wide variety of tasks that have been used to investigate this fundamental aspect of cognition. Typical tasks in the literature of interval timing concern intervals that subjects have previously experienced between a sequence of events. Some require the use of motor skills (e.g. via the reproduction of an interval between sensory cues or actions), where motor timing is involved \citep{Paton:2018}, while some tasks focus only on the perception of these intervals (i.e. sensory timing tasks). In the latter category, subjects can discriminate intervals in a variety of ways, such as by estimating the elapsed time in human-readable units, categorization, bisection, or by comparison between multiple intervals. Finally, in all cases, cues can involve individual sensory modalities (e.g. visual or auditory) or a combination of more than one \citep{Ellinghaus:2021}.

These rigorous and highly simplified experimental settings have led psychophysical studies to unravel consistent patterns of bias that the brain exhibits in different statistical contexts. Since the duration of an interval is a magnitude, it is not surprising that the most well-established biases include the contextual effects observed in other types of magnitude judgements, such as loudness or distance \citep{Petzschner:2015}. First, subjects have an overall tendency to regress duration judgements to the mean duration of stimuli already presented to them, which is referred to as the \textit{central-tendency} effect \citep{Jazayeri:2010}, or \textit{Vierordt's law}, after the German physiologist Karl von \cite{Vierordt:1868} who first presented evidence for this effect. This central tendency is naturally present in different timing tasks, timescales, sensory modalities and age groups \cite{Lejeune:2009}, although evidence suggests that it can be alleviated via training (e.g. in expert musicians \citep{Cicchini:2012}). Second, as in other dimensions of stimulus perception, the ability to perceive changes in duration is characterized by being directly proportional to the actual duration \citep{Gibbon1977ScalarET}. This is known as the scalar property of interval timing, or Weber's law.

In addition, it is shown that the order in which two stimuli of variable durations are presented determines the reported duration judgements \citep{Dyjas:2012}. This is due to the `local' contextual bias, as opposed to the `global' bias captured by Vierordt's law, which is manifested as a tendency of current estimates to be pulled towards the preceding stimuli \citep{Dyjas:2012,Shi:2013,DeJong:2021}.
Finally, other contextual effects in interval timing include deviations in duration judgements when subjects experience increases in cognitive load \citep{Block:2010}, as well as distinct biases across different scene types when subjects watch videos \citep{Roseboom:2019}.

The first step in understanding the nature of these contextual biases is to identify the sources of information that the brain uses to process durations. This is less straightforward than other types of measurable dimensions of perception, such as the height of an object, since duration constitutes an intrinsic property of experience that does not necessarily depend on any particular cue or sensory modality. Nonetheless, this very omnipresence shows the wide repertoire of information sources that the brain is able to employ in order to improve its performance in interval timing tasks. 
Following Occam's razor principle \citep{Feldman:2016}, the brain should learn to exploit the least complex and reliable source for a given task and range of durations. For instance, when trying to accurately estimate a blank interval of up to a few seconds in a simple experimental setting, it is reasonable to assume that the brain would employ any kind of neural dynamics that exhibit periodicity (such as cortical oscillations) or other rhythmic physiological signals (such as the heartbeat) as a frame of reference. Indeed, evidence shows that, in simple tasks, information related to the cardiac cycle is relevant for perception and reproduction of time intervals in the range of 2-25 seconds \citep{Meissner:2011,Pollatos:2014}, while it has been shown that other interoceptive signals (such as hunger -- see \cite{Vicario:2019}), might also play an important role.

On the other hand, estimations of duration in everyday phenomenological experience are more reasonable to distil evidence predominantly from exteroceptive signals, which provide a far richer source of temporal information \citep{Ahrens:2011}. This was examined by \cite{Suarez:2019}, who showed that supra-second duration judgements vary systematically with perceptual content, rather than physiological signals, in an experiment where participants are viewing natural videos.

Over the years, a substantial number of different theoretical proposals have attempted to explain human behaviour in interval timing tasks. Although a comprehensive taxonomy of interval timing models is not in the scope of this chapter, for an extensive overview of theoretical models the reader is directed to the existing reviews on this topic, such as \cite{Addyman:2016} or \cite{Basgol:2021}.

Perhaps the most influential type of such models involves a pacemaker and an accumulator which together constitute the main components of an \textit{internal clock} mechanism \citep{VanRijn:2014}. Formally proposed by \cite{Gibbon1977ScalarET} as scalar expectancy (or timing) theory in order to model behaviour governed by time, this internal clock can be viewed as a part of a larger cognitive architecture involving a clock, memory and decision stages \citep{Treisman1963TemporalDA,Church1984PropertiesOT}. In 2004, \cite{Matell:2004} proposed the striatal beat-frequency model, a neurobiologically plausible implementation of the clock stage in scalar timing theory \citep{VanRijn:2014}, which relies on oscillatory properties of cortical neurons that project to the basal ganglia to create a pulse. In addition, in a leading alternative theory of interval timing, \cite{Karmarkar:2007} proposed that time can be tracked by stochastic neural processing dynamics within any given state-dependent network, without the need for dedicated clocks.

\subsection{Bayesian approaches to interval timing}

The aforementioned context effects in duration estimation point to the existence of a Bayesian mechanism for the integration of representations in memory and new sensory information, in line with other types of magnitude \citep{Petzschner:2015}. Based on this framework, previously acquired (prior) information of estimated intervals is integrated with noisy sensory input (likelihood), weighted by their relative uncertainty, to produce a statistically optimal posterior estimation. Consequently, the central tendency effect can be explained as a bias towards prior expectations, in cases of disagreement between memory and current physical stimulus.

\begin{figure}
 \centering
    \includegraphics[width=\linewidth]{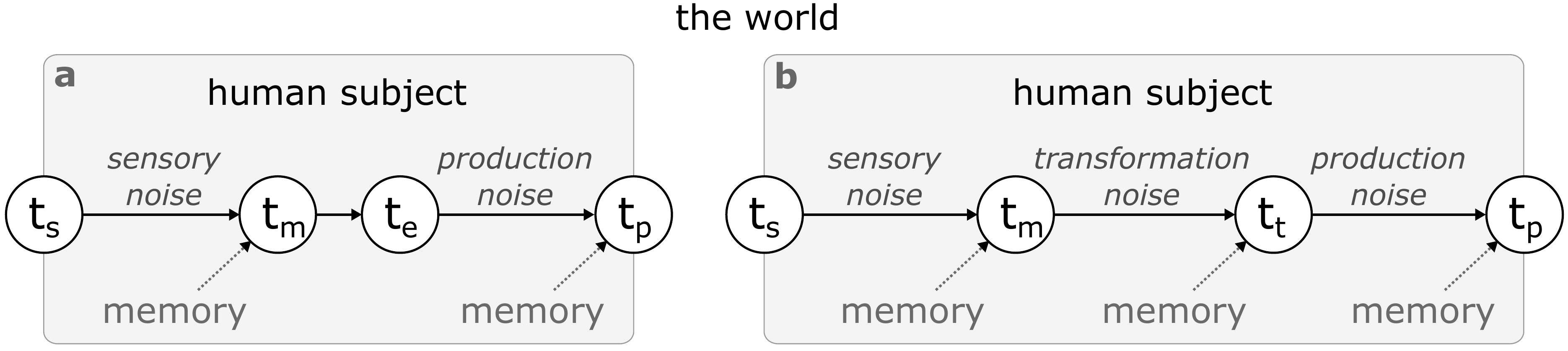}
    \caption{Bayesian observer-actor model by \cite{Jazayeri:2010} (\textbf{a}) and a more general version by \cite{Remington:2018} (\textbf{b}). The graph shows causal relationships between random variables. Inference involves integrating new information (horizontal arrows) with the corresponding prior statistics found in memory. In (a), $t_s$ represents the set of external stimuli, $t_m$ the likelihood distribution of measured duration of the event, given the memory contents, $t_e$ the result of applying to the later distribution a Bayes estimator, and $t_p$ the reproduced duration. In (b), $t_t$ represent the distribution of the estimated duration, taking into account more noise steps in the computation.}
    \label{fig:BOM}
\end{figure}

This idea was explored by \cite{Jazayeri:2010}, who proposed the use of a \textit{Bayesian observer} to model human perception and reported time estimates. In their model, also shown in Figure~\ref{fig:BOM}a, the process of inferring the measured duration ($t_m$) given an external sample ($t_s$) is performed using a Bayes estimator (outputting $t_e$). The likelihood of a measurement is expressed as a probability distribution that parametrises any noise originating from the process of sensing via different modalities, while the prior is set to be a fixed uniform distribution within the range of all earlier observed durations. Then, the reproduced duration by human subjects ($t_p$) is expressed as a Normal distribution centred at $t_e$ and with a standard deviation that captures any motor noise. Importantly, both sensory and motor noise factors in this model were conceptualized to grow linearly with the corresponding means, a property that was crucial to capture Weber's law. This model was shown to fit well with the behaviour of subjects that are tasked to reproduce durations sampled by different distributions matching the model priors, especially when the estimation of the posterior duration is performed using the Bayes least-squares method.

\cite{Jazayeri:2010} also formulated a \textit{Bayesian observer-actor} model where the process involves optimization of all steps: from sensing a duration to motor reproduction. This model accounts for uncertainty (or noise) minimization at all stages of a time interval task, including motor control. Furthermore, a general Bayesian observer-actor model might involve more complex transformations of time measurements to behaviourally meaningful variables. Such transformations can be viewed as any kind of downstream computation that is necessary for the accomplishment of a task using general representations of time and can be expressed as a mapping $t_m \rightarrow t_t$ (Figure~\ref{fig:BOM}b). For instance, when asked to verbally report the duration of a video, human subjects have to map their inner sense of time to units that can be expressed in a human language, such as seconds. Similarly, when asked to reproduce the same duration by pressing a button twice, subjects have to map the same inner sense to a representation that is relevant to the dynamics of the motor system, related to the part of the body that controls the button. Finally, in the case of reporting durations using a visual analogue scale, it is necessary to also take into account the visual perception of space. Arguably, the three types of transformations involved in our examples would produce different levels of \textit{expected} noise, whose effect, in turn, would require different sets of skills from memory to be mitigated.

In a later iteration of this Bayesian observer-actor model for interval timing, \cite{Remington:2018} investigated these aspects and suggested that the brain is not only able to apply inference in mental transformations $t_m \rightarrow t_e$, but also to even `\textit{delay}' early-stage inference in sensory measurements $t_s \rightarrow t_m$ in order to account for both types of expected noise (sensory and transformation) together.

Although the basic form of the general Bayesian observer-actor model is very useful as a backbone architecture of the computations involved in magnitude perception, specifications of individual components are often defined arbitrarily, causing substantial criticism \citep{Jones:2011,Bowers:2012,Wei:2015}. Hence, it is worth considering some individual components used in interval timing models in more depth. Two of the most crucial components are the likelihood and the prior of the sensory measurements. The likelihood, modelled as simple Gaussian noise in the studies shown in Figure~\ref{fig:BOM}, can be thought of as the basic `sense' of time that is combined with new sensory information from different sources, as mentioned in the previous section. Depending on the sources taken into account, studies incorporate completely different versions of parametrised probability distributions to model likelihood.

In addition, there is a wide variety of proposals for modelling prior statistical knowledge as a probability distribution of intervals. Notable examples of densities include Uniform \citep{Jazayeri:2010}, Gaussian \citep{Cicchini:2012}, non-parametric \citep{Acerbi:2012} and, most recently, a mixture of lognormals \citep{Maas:2021}. These densities are either considered fixed across experiment trials or dynamic, with parameters being updated using a learning rule. The mechanism used for updating memory representations is a crucial aspect of Bayesian observer models and it determines whether or not both global and local contextual biases can be captured. \cite{DeJong:2021} recently showed that a linear recursive Bayesian estimator, in particular a Kalman filter, is better at capturing human global and local biases than the alternative non-Bayesian approaches. This points to the existence of a single dynamic representation of prior information in the brain that is responsible for both types of biases. Corroborating this view, \cite{Ellinghaus:2021} recently showed evidence indicating that a general dynamic prior for duration judgements integrates temporal information from different sensory modalities.

Finally, although the basic sense of time is more closely related to individual measurements, it can be argued that at least some aspects of the human experience correspond to the inferred posterior durations that integrate multi-modal information with memory \citep{Helmholtz:1867,Seth:2015,Neemeh:2020}. In particular, studies of the phenomenology of time have suggested a separation that comprises \textit{implicit} (or lived) and \textit{explicit} time \citep{Fuchs:2013}. The former describes the experience of `being inside' time since it is associated with self-awareness and intentionality and requires a continuous flow of experience. In contrast, explicit time refers to the measurable time that we can process in the past, present and future. According to Fuchs, explicit time emerges when the flow of implicit temporality is interrupted. The difference in the types of temporal experience is further illustrated by \cite{Droit:2016}, who showed that the perception of the passage of time is mediated through a different mechanism than the perception of duration.

\section{Sense of time as perceptual change}

A plethora of both classical philosophical work \citep{Selby:1964} and behavioural studies \citep{Block:1974,Ornstein:1975,Poynter:1983} suggest that the perception of durations is driven by changes in the contents of sensory experiences. This idea is in contrast to predominant approaches mentioned before that rely on the existence of rhythmic neural processes to act as pacemakers \citep{Treisman1963TemporalDA,Matell:2004,VanRijn:2014}. Recently, \cite{Roseboom:2019} proposed a computational model that implements this idea using the popular large-scale artificial neural network AlexNet by \cite{Krizhevsky:2012}. 

The connectivity of this neural network is organized using convolutions which are shown to result in similar hierarchical structure and representations as in the primate cortex, including regions associated with vision \citep{Cadena:2019,Kriegeskorte:2015,Khaligh:2014}. AlexNet is trained to classify static, high-resolution images into 1000 different categories of everyday objects and animals. In the model of \cite{Roseboom:2019}, AlexNet received consecutive frames of videos in order for its neuron values to be analogous to biological neuron states and reflect dynamic latent representations in the human visual perception system. To capture salient perceptual changes, the Euclidean distance between consecutive neuron states across the network's hierarchy was calculated and compared to a threshold value, resulting in a salient event detection mechanism. The events in a video were then accumulated and used as the inner sense of time, similar to the measurement $t_m$ in Figure~\ref{fig:BOM}b. Finally, a regression algorithm was used to map accumulated events to seconds, corresponding to the mental transformation $t_t$ of the same figure.

Interestingly, this computational model was found to fit well the behavioural data of humans when judging the durations of videos of natural scenes that lasted between 1 to 64 seconds, exhibiting crucial patterns of bias examined before, such as the Vierordt's and Weber's laws among others. Indeed, this was despite the fact that the mapping $t_m \rightarrow t_t$ was trained to capture the objective duration of the videos, as well the fact that the neural architecture was trained on a completely different domain and purpose than this task.

More recently, \cite{Sherman:2020} ran the same experiment as \cite{Roseboom:2019} while recording the subject's brain activity using Functional magnetic resonance imaging (fMRI) and further verified the core suggestions of the original study. Analysis of these recordings showed that blood flow changes in the visual cortex were sufficient to reconstruct trial-by-trial human biases, while this was not the case in other perceptual regions of the cortex that were not involved in processing the video stimulus.

Although the theoretical model by Roseboom and colleagues is not defined as a Bayesian observer model, it is clear that it follows the same principles and could be adapted to the backbone architecture by \cite{Remington:2018}.
Importantly, it shows that there is no need for a Gaussian noise component in the measurement likelihood of these models and that increasing the noise arbitrarily as a linear function of the means can be replaced by accumulated salient changes, without the need of any parameters except for a pre-trained perceptual system, while maintaining consistency with Weber's law.

\subsection{Time, predictive coding and episodic memory}

As pointed out in \cite{Roseboom:2019}, the Euclidean distance between successive neuron activations in the model of this study can be viewed as a proxy for the error between the prior prediction of latent states and the corresponding posterior. This suggests a hierarchical architecture where representations of all neuron layers are predicted and updated, via the process of probabilistic inference. This approach to perception is known as predictive coding and was popularised as a computational model of perception by \cite{Rao:1999}.

There are multiple associations between this framework and the perception of time. \cite{Hohwy:2016} proposed that the hierarchical nature of predictive coding is what creates the phenomenology of temporal flow. While the brain's model of the environment predicts changes in the current state of affairs, it begins distrusting the perceived present in anticipation of these changes, thus creating the illusion that the present moves forward. \cite{Kent:2019} extended this theory and proposed that distrusting the future can explain the severe experience of time dilation in depression and hopelessness.

Finally, predictive coding has been used to explain the neural mechanism that underlies the perception of duration. In a recent study by \cite{Fountas:2021} and based on the same fundamental assumptions as  \cite{Roseboom:2019}, models of attention, episodic memory and visual perceptual processing were integrated under the framework of predictive coding. In a large-scale experiment involving nearly 13,000 participants, the resulting system was able to reconstruct key human biases in duration judgements regardless of the stimulus content, whether the participants focused on this task or they were distracted, or even when judgements were requested retrospectively, forcing participants to rely solely on memory recall. The success of this model was crucial since prospective and retrospective human duration judgements are known to show inverse patterns of bias \citep{Block:2010}, which are notoriously hard to capture together in a single theoretical framework \citep{Basgol:2021}.

\section{Time and Artificial Intelligence}

The ability to perceive and estimate time may well be one of the central elements of an intelligent biological agent equipped with a model of its environment. Indeed, it would be of limited use for an organism's survival to learn the causal physics of the world without its underlying temporal dynamics and structure. As such, if one takes the view that an agent's internal generative model of the environment is its primary guide to behaviour, the ability to learn representations of time and employ them for action selection is an important consideration for building intelligent artificial agents. In this section, we wish to consider time perception as it pertains to artificial intelligence and from the perspective of a researcher whose primary objective is to create agents capable of intelligent interaction with an environment.

\subsection{Model-based reinforcement learning}

\subsubsection{Overview}

Within the field of artificial intelligence, a large community of researchers are focused on studying reinforcement learning (RL) -- a research area concerned with how agents ought to behave to maximise rewards in an environment. Since the wide adoption of \textit{deep learning}, popularised by the success of DeepMind \citep{Mnih2013PlayingAW, Mnih2015HumanlevelCT}, RL has been in a time of renaissance with an increasing number of works demonstrating its convincing efficacy. 

Today, a large number of agent architectures have been proposed that were shown to solve complex tasks in a wide variety of environments \citep{Silver2014DeterministicPG, Schulman2015TrustRP, Lillicrap2016ContinuousCW, Mnih2016AsynchronousMF, Blundell2016ModelFreeEC, Racanire2017ImaginationAugmentedAF, Ha2018RecurrentWM,  Haarnoja2018SoftAO}. Crucially, current prominent RL architectures are able to solve many tasks involving temporal dependencies and interval timing \citep{LongTermAndMemory, MERLIN, HumanLevel-DeepMind, ArjonaMedina2019RUDDERRD} but still struggle when it comes to long-term planning and credit assignment. The challenge of temporal learning is a growing and important area of research in the field, with a simple question underlying its foundations: do agents learn time? 

Such questions, however, are bound to fail given their intrinsic equivocacy. Similar to the study of time in humans, \textit{learning time} in RL can be viewed from different lenses given the grand scale of problems and research branches this field encapsulates. In particular, it can be viewed unambiguously as the ability to estimate time (interval timing) \citep{Deverett:2019}, as the ability to learn the temporal dynamics of the environment (transition function) \citep{Sutton1990IntegratedAF, 01-09=E2C, 01-12=PILCO, WorldModels}, or as the ability to perform accurate credit assignment to events substantially separated in the temporal domain \citep{LongTermAndMemory, KeLongTerm}. As such, a better question to ask is \textit{how} do the currently proposed RL architectures learn time and how do these methods fit into the conception of a Bayesian brain capable of representing it?

To address these questions, it is important to acknowledge that RL is commonly divided into two categories that relate to the two basic approaches to building agents -- model-free and model-based. 

\begin{wrapfigure}{r}{0.5\textwidth}
 \centering
    \includegraphics[width=0.95\linewidth]{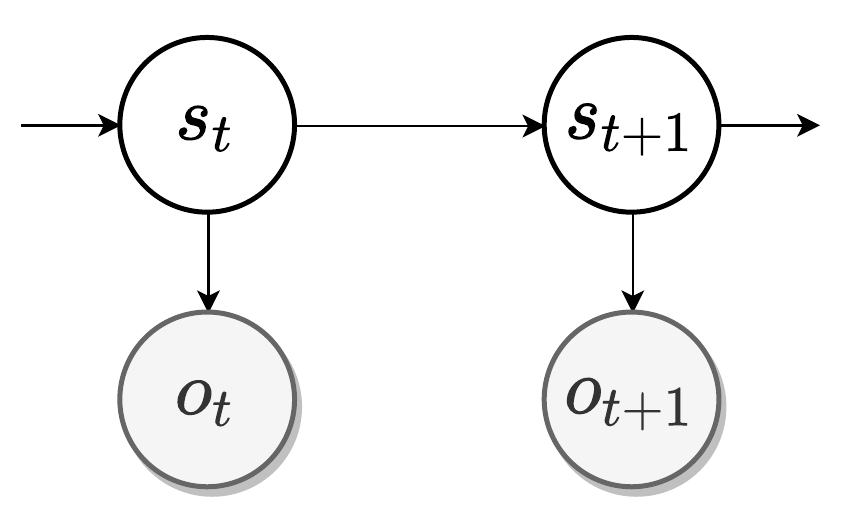}
    \caption{Graphical model of a partially observable Markov decision processes (POMDP) generative model. Latent state $s_t$ is the current state of the environment at time $t$ with the corresponding observation $o_t$ available to the agent.}
    \label{fig:pomdp}
\end{wrapfigure}

The model-free approach encompasses a range of well-known methods \citep{Mnih2015HumanlevelCT, Silver2014DeterministicPG,  Schulman2015TrustRP, Lillicrap2016ContinuousCW, Mnih2016AsynchronousMF,  Haarnoja2018SoftAO} and involves learning of a mapping between states (or observations) and actions without constructing an internal model of the environment. Most commonly, model-free algorithms rely on the idea of a value function -- the expected cumulative reward associated with a state of an environment -- and act according to maximising the action-value pair in any given state. 
On the other hand, model-based RL comprises methods that equip an agent with a learnable model of its environment, which can be used for deliberate action selection by means of planning. Importantly, this category of agents is conceptually closer to the biological intelligence realised in humans, given that the contemporary view of the brain is that of a Bayesian generative model. It is for this reason that we will focus our discussion on model-based RL.

Most of the model-based RL agents are Bayesian, in a sense that they possess probabilistic generative models and update their beliefs via Bayes' rule. These generative models are most commonly formulated as partially observable Markov decision processes (POMDP) -- a generalisation of MDP \citep{Bellman1957AMD} for environments where the true state cannot be directly observed (Fig.~\ref{fig:pomdp})\footnote{Please note that we exclude actions from the formulation of a generative model to be more consistent with the models discussed in Section \ref{sec: generative_models}. However, for RL, actions are naturally required for defining a model. For instance, a transition dynamics model is typically written as $p(s_{t+1}|s_t, a_t)$, where $a_t$ is an action made by an agent at timestep $t$. Nevertheless, we thought it would be superfluous to incorporate actions into the discussion focused on the generative models underlying agents' perception and temporal learning.}. In such environments, agents hold \textit{beliefs} over states, $p(s)$, which are iteratively updated upon receiving new information in a form of environment observables, $o$. Importantly, as the agent interacts with the environment, its state changes over time, as do the agent's beliefs. This \textit{belief} dynamics is typically modelled by an explicit component called a \textit{transition dynamics} function, expressed as $p(s_{t+1}|s_t)$ (or $p(s_{t+1}|s_t, a_t)$ including actions -- see footnote 1). As such, in its simplest form, an internal model of an agent constitutes a set of beliefs it holds about the state of the world (inferences) and a parametrised transition function modelling the dynamics of its own beliefs (and by consequence of the environment).

As we discuss the state of the art in Bayesian model-based RL, we want to draw the reader's attention to a plausible interpretation of what a generative model represents with respect to temporal learning. Apart from its unambiguous function of being an agent's computational cognitive model, the contents (i.e. learned representations of the world) of a generative model tell a story about the agent's \textit{subjective} experiences of the environment. In particular, the emergent properties implicit into the components of its generative model can be viewed as the agent's means of perceiving and estimating time, and are therefore of great interest for the community. 

\subsubsection{Time and internal models}
\label{model_based_rl}

\begin{figure}[h!]
 \centering
    \includegraphics[width=0.8\linewidth]{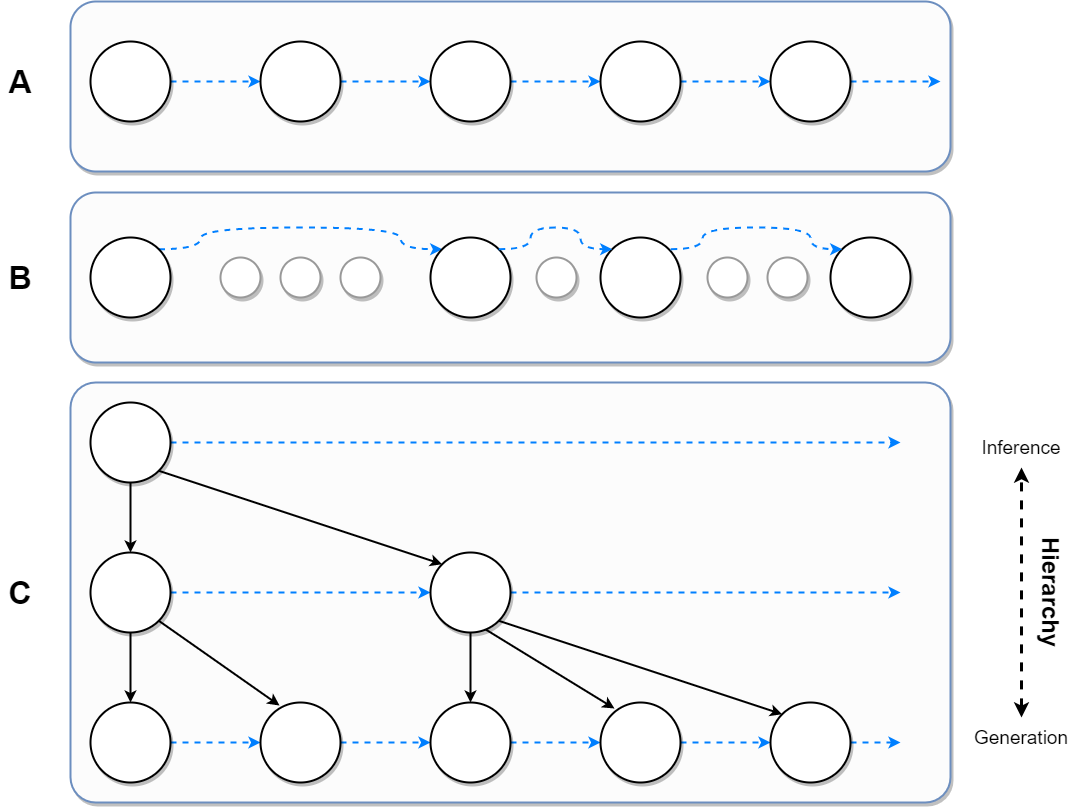} 
    \caption{Types of generative models for temporal learning in the field of AI. Circles represent latent belief states, while arrows represent the generative process. \textbf{A.} Traditional Markovian transition models. \textbf{B.} Variable-timescale transition models that transition over a distinct timescale as opposed to the physical timescale. \textbf{C.} Hierarchical generative models with nested timescales.}
    \label{fig:generative_models}
\end{figure}

\paragraph{Traditional models} Earlier works on deep model-based RL paid little consideration to learning temporal properties, focusing instead on the basic Markovian (or feed-forward) transition models \citep{Sutton1990IntegratedAF, 01-12=PILCO, 01-02, 01-03}. Markovian transition models are trained to map a belief state to its temporal successor, $p(s_{t+1} | s_t)$, but not beyond, adhering strictly to the POMDP structure (see Fig.~\ref{fig:generative_models}A). As such, they are myopic and hold no explicit internal information about the past. In considering the consequences of their myopic nature, it becomes evident that feed-forward agents have little access to the temporal information of the environment and must therefore rely on other sources for their representation and estimation. Indeed, \cite{Deverett:2019} demonstrated that feed-forward agents gather temporal information from the environment, exhibiting autostigmergic behaviour\footnote{Autostigmergic behaviour refers to situations in which an agent's actions result in the formation of traces in the environment that is perceived by the agent in the future as an incentive to further develop the action undertaken.} in which the agent's interactions with the environment are used as an external source of memory about the elapsed time. Similarly, the agent's perception of the environment's temporal dynamics is formed exclusively by the pairs of the inferred states and the change that separates them in the belief space. Despite the fact that Markovian models have generally been shown to be inferior in designing agent generative models, \cite{Deverett:2019} suggest that the method by which the temporal properties of the environment are gathered and the emergent estimation biases demonstrated in the work (scalar variability and mean-directed bias at the tails) resemble those found in primate experiments. 

More recent and prominent model-based architectures demonstrate superior performance to their earlier feed-forward counterparts with the use of autoregressive (recurrent) transition models (deep learning models such as LSTM \citep{LSTM} or GRU \citep{GRU}) \citep{Dreamer, Dreamer2, Schrittwieser2020MasteringAG}. For instance, the Dreamer agent \citep{Dreamer} is equipped with a recurrent state-space model (RSSM), in which the transition dynamics is an autoregressive model that incorporates internal information about previously processed states, $p(s_{t+1}|s_{\leq t})$. Notice that compared to a feed-forward transition model, an autoregressive model predicts the next timestep $s_{t+1}$ conditioned on all previously inferred states of the world $s_{\leq t}$. This causal structure of the model allows it to possess information about the elapsed time in a form of a clockwork, stored in the patterns of its neural activations \citep{Deverett:2019}. As such, it has been demonstrated that recurrent models are not subject to the biases found in time estimation experiments with humans and are computationally superior by virtue of being capable to use more information about the past. 

Despite the apparent effectiveness of the latest autoregressive agents driven by the idea of attaining more rewards, the ideas standing behind their design touch little upon temporal learning. For instance, consider again the two kinds of transition models (feed-forward and autoregressive), as well as the definition of a POMDP, we presented earlier. A key indexing unit involved is that of the \textit{physical time} of the environment, $t$, referring to the objective succession of time -- independent of any one observer. Could modelling the objective timescale of the environment be called subjective perception of time? Does time perception remain intact regardless of the context? And, more practically, is the objective timescale optimal for planning and action selection?

\paragraph{Variable-timescale models} It seems that these questions, at least in part, have motivated a number of works that went beyond the classical definitions of internal models and considered the \textit{timescale} as an indirectly learnable component -- an ambition which may be viewed as an unintentional attempt to build computational models of time perception in RL. Indeed, the common overarching theme of these works is in the defining of a more suitable timescale for agent transition models (see Fig.~\ref{fig:generative_models}B). The methods generally differ in the criteria by which the timescales are determined and may be viewed as the distinct approaches to the implementation of time perception in computational cognitive models. 

For instance, \cite{Neitz:2018} and \cite{Jayaraman:2018:TAP} define a model that learns transitions between most predictable states. The resultant property of such a model is the ability to predict the future (and thus perform planning) with respect to the most anticipated events in a potential sequence. Although these models are not probabilistic, their mechanisms can be viewed from the Bayesian perspective -- as taking a prediction path of least belief update. Specifically, under this approach, a probabilistic transition model, $p_\theta$, is optimal if its parameters minimise the belief update between the model's prediction from the current timestep $t$ and some future timestep $k$,
\begin{align}
    \theta^{*} = \arg \min_{\theta} \min_{k \in T} D_{KL} \big[ \text{ } p(s_{k}|o_k) \text{ } || \text{ } p_\theta(\bar{s}|s_t) \text{ } \big],
\end{align}
where $p(s_k|o_k)$ is a posterior latent state at some timestep $k \in T$ that minimises the Kullback-Leibler divergence under the predicted prior distribution, $p_\theta(\bar{s}|s_t)$, and $T$ is a set of timesteps $>t$. Recall similarly that the presented quantity is also often called Bayesian surprise, more generally written as $D_{KL} \big[q||p\big]$. As such, predictability criterion used in the two works, relate directly to this quantity when translated into the language of Bayesian belief updating, and correspond to its minimisation in the characterisation of a timescale. Such interpretation of the approach is important as it allows us for a direct comparison with other works in the Bayesian framework. 

In particular, \cite{Zakharov:2021} introduced \textit{subjective-timescale} models (STMs) that similarly rely on the Bayesian surprise for defining the timescale an agent would be trained to reproduce. However, unlike the approaches by \cite{Neitz:2018} and \cite{Jayaraman:2018:TAP}, STMs select the states if the Bayesian surprise \textit{exceeds} a certain threshold -- a principally different approach in which a model tends to take the prediction path of \textit{most} belief update,
\begin{align}
\label{eq2}
    \theta^* = \arg \min_{\theta} D_{KL} \big[ \text{ } p(s_{k}|o_k) \text{ } \| \text{ }  p_\theta(\bar{s}|s_{t}) \text{ }  \big],
\end{align}
for $k$ iff,  
\begin{align}
\label{eq3}
    D_{KL} \big[ \text{ } p(s_{k}|o_k) \text{ } \| \text{ }  p_{\psi}(s_k|s_{k-1}) \text{ }  \big] > \epsilon,
\end{align}
where $\epsilon$ is a Bayesian surprise threshold. Here, STM ($p_\theta$) is trained to transition a belief state to a future state that caused inequality in Eq.~\ref{eq3} to be satisfied. Notably, Bayesian surprise in Eq.~\ref{eq3} is computed using the physical-timescale transition model, $p_{\psi}(s_{k}|s_{k-1})$. The authors hypothesise that the surprise quantity represents the visual salience of the agent's sensory inputs \citep{Zakharov:2021}. This work speaks to the ideas from an earlier discussed computational model of time perception that views saliency (Bayesian surprise) as the basic metric underlying the subjective experience of time \citep{Fountas:2021}. More saliency means the distance between states in a physical timescale reduces, while its absence stretches this distance out. By using a number of perceptually complex datasets, \cite{Zakharov:2021} demonstrated the presence of this emergent property in a generative model of a trained RL agent. Similarly, when applied to a well-established model-based agent, Dreamer \citep{Dreamer}, STM significantly improved its performance thus indicating its practical utility for planning and action selection.

Other works have concentrated on the `informativeness' of states in defining the timescales. \cite{Pertsch:2020} introduce a two-level hierarchical generative model that learns to transition between the so-called `keyframes' of a video sequence in the top level of its hierarchy (with bottom level states being conditioned on pairs of the keyframes above). The keyframes are picked out by a parametrised stochastic model and therefore define the relevant timescale of a video sequence. Importantly, the keyframing model is a component of an agent's generative model and is thus trained in conjunction with the rest of the model by means of minimising the joint lower bound on the marginal likelihood. The informativeness of the states is considered implicitly -- if a particular predicted keyframe is useful for the maximisation of the likelihood of a predicted sequence, it is more likely to be picked as a keyframe. 

This work represents a fundamentally different approach to defining a more useful \textit{subjective timescale} for an agent perceiving its environment. First, the integration of a separate timescale-constructing model into the agent's generative model draws an interesting connection with biological time perception hypotheses on internal clock and dedicated models \citep{Treisman1963TemporalDA, Gibbon1977ScalarET, Church1984PropertiesOT}. In particular, this method considers the existence of a separate model for determining the rate at which the subjective time flows (by means of keyframes). Second, it contrasts with the previously discussed approaches where timescales are defined using non-parametric techniques and which rely solely on the reactive properties of the agent's generative model. There, the timescale-defining quantity is a form of Bayesian surprise, computed by the synthesis of sensory input with a prediction made by an agent's internal model -- a technique that may be viewed to be more in line with predictive coding.  

\subsection{Hierarchical generative models}
\label{sec: generative_models}

As explained, model-based RL is a particularly attractive area for exploring computational models of temporal learning given its conceptual similarity to the contemporary view of the brain as a statistical machine. Furthermore, it deals with agency and thus a perception-action loop that ultimately results in a unique sequence of observations and internal representations experienced and formed by an agent. However, the field of RL, though intricately connected to the areas of representation learning and generative modelling, has not fully explored some of the more advanced research performed in these areas. One such instance is that of \textit{hierarchical} generative models. The role of hierarchical processing in the brain has been demonstrated in numerous works through the years, including hierarchical processing of time. In particular, empirical evidence from neuroimaging studies suggests that the human brain employs its hierarchy to segment continuous data (deducing its discretised structure in a form of event sequences). Across the nested hierarchy of its cortex, the brain represents timescales ranging from milliseconds to minutes \citep{Baldassano:2017}, with an increasing level of temporal abstraction along its computational layers. This makes the branch of hierarchical generative modelling very attractive in studying how models, akin to our brains, could be built to exhibit similar properties of temporal learning and abstraction.  

Previously, we have discussed one such work by \cite{Pertsch:2020}, in which a two-level hierarchical model has been employed to reconstruct video sequences with the use of a parametrised keyframe detector. Another work by \cite{Kim:2019:VTA} incorporates a similar technique into its two-level generative model. Although the formulation of the graphical model in \cite{Kim:2019:VTA} is different to \cite{Pertsch:2020}, in both models the physical-timescale variables are conditioned on the second level `keyframes' (or \textit{boundary} states). A simplified version of a temporal hierarchical generative model, in which lower-level states are conditioned on the higher-level states is shown in Figure \ref{fig:generative_models}C. Here, it is pertinent to ask: apart from the aforementioned biological plausibility, what is the practical and conceptual merit of hierarchies, particularly for temporal learning?

The temporal hierarchy of latent variables force a generative model to exhibit certain properties that would otherwise be difficult, and perhaps even impossible, to acquire. First, nested hierarchies (such as the one shown in Fig.~\ref{fig:generative_models}C) naturally result in the \textit{increasing temporal abstraction} across layers, as higher-level states correspond to a larger number of physical-timescale states and are employed during the generative process to produce lower-level states. Similarly, higher-level states are inferred less often, promoting the representation of increasingly slower spatiotemporal features in those levels. Second, hierarchies lead to \textit{segmentation} of sequential data by means of higher-level latent variables (also called \textit{contexts}) that govern the generation of the states below and serve as boundaries between lower-level sequences. As discussed, segmentation of events is an evident component of the human brain's perception machinery \citep{hard2006making, kurby2008segmentation}, making computational cognitive models with this capability important for understanding how it could be realised in the brain. 

Still, generative models with nested hierarchies present a number of practical challenges since it is unclear as to how they should be constructed. In particular, when should the contexts be updated and how could this be realised within a Bayesian model? Perhaps for this reason, a number of works have proposed fixed-interval models, in which each hierarchical level of a model is assigned its own manually-defined rate of update \citep{lotter2016deep, Koutnik:2014:CW-RNN, Saxena:2021:CWVAEs}. Although fixed-interval models are capable of distributing abstract spatiotemporal features across its levels, as shown in \cite{Saxena:2021:CWVAEs}, the emergent representations are not satisfactory with respect to the temporal factors of variation in sequential data \citep{Zakharov2021VariationalPR}. In particular, event segmentation, and thus an optimal discretisation of a sequence, is simply not possible within this framework, as the temporal structure of the generative model is pre-determined regardless of the incoming sensory input. Furthermore, this leaves little room for the presence of \textit{subjective} time perception in a manner presented previously \citep{Zakharov:2021} -- a phenomenon that has been observed in human experiments.

Accordingly, some authors proposed models with adaptive (flexible) timescales across the hierarchical levels of a generative model. \cite{Chung:2017} introduced Hierarchical Multiscale Recurrent Neural Networks (HM-RNNs) that employed a discrete-space long short term memory (LSTM) model with a parametrised boundary detector. The authors demonstrated that a hierarchical LSTM can discover an underlying spatiotemporal structure in sequential data, facilitating further research into this area. Notably, HM-RNNs were not probabilistic models and therefore did not operate over Bayesian \textit{belief} states. This led \cite{Kim:2019:VTA} to extend HM-RNNs to stochastic models (though only possessing two hierarchical levels) with a conceptually analogous approach of accommodating a parametrised boundary detector as part of the generative model. 

More recently, \cite{Zakharov2021VariationalPR} presented a Variational Predictive Routing model (VPR) that employs a non-parametric event boundary detection mechanism at every level of its latent hierarchy. The detection mechanism is primarily used to enforce a structure of \textit{subjective} nested timescales that update only with respect to the features represented in their respective layers. As evident from the results demonstrated in the work, VPR has an elegant interpretation with respect to the potential cognitive mechanisms involved in its functionality and is hypothesised by the authors to be an effective time perception model. 

VPR's boundary detection system employs Bayesian belief update quantities to decide on whether a hierarchical state should be updated. Specifically, the authors claim that the model is capable of capturing expected (predictable) and unexpected (surprising) events in every level of its hierarchy using two criteria. Upon detecting an event at level $n$, the model replaces its belief state $s^n$ with a newly inferred posterior -- resulting in a structure of a generative model similar to the one shown in Figure \ref{fig:generative_models}C. Notice the nested structure of the graphical model, in which the higher-level states are allowed to update only if all lower-level states have similarly been updated. 

The decision to infer a new hierarchical posterior state is decided using the mentioned event detector. Unexpected events are identified by considering the inequality,
\begin{align}
    D_{KL}(q_{st} || p_{st}) > \epsilon,
\end{align}
where $p_{st} = p(s_{\tau+1} | o_\tau, s_{<\tau})$ is the model's prior over the next belief state signifying the assumption that the belief will not change (\textit{static assumption}), $q_{st} = q(s_{\tau+1} | o_{\tau+1}, s_{<\tau})$ is the inferred posterior given the new observation $o_{\tau+1}$, and $\epsilon$ denotes a threshold beyond which an unexpected event is regarded as detected.  For simplicity, the hierarchical levels of belief states are not included in the notation.

Notice that under the so-called static assumption, the state of the generative model remains intact (i.e. the latent variables employed during inference, $s_{<\tau}$, stay constant), allowing for a clear interpretation of the divergence. In particular, the two distributions are identical if the observations have not changed over time, $p_{st} = q_{st}$ iff $o_\tau = o_{\tau+1}$. As such, this Bayesian surprise quantity can be interpreted as a measure of how much the model's beliefs about the encoded features have changed over time. The threshold $\epsilon$ thus serves to identify changes that are \textit{sufficiently} significant to be classified as an event. 

The second criterion employed by VPR is for the detection of expected events, 
\begin{align}
    D_{KL}(q_{st} || p_{st}) > D_{KL}(q_{ch} || p_{ch}),
\end{align}
where $p_{ch} = p(s_{\tau+1} | s_{\tau}, s_{<\tau})$ is the transition model's  prior belief over the state it expects to observe at the next timestep (hence, it is the \textit{change assumption}), and $q_{ch} = q(s_{\tau+1} | o_{\tau+1}, s_{\tau}, s_{<\tau})$ is the inferred posterior state given a new observation $o_{\tau+1}$. It is similarly intuitive to interpret this quantity: if a prediction made by the model, $p(s_{\tau+1} | s_{\tau}, s_{<\tau})$, is consistent with a newly inferred posterior state, $q(s_{\tau+1} | o_{\tau+1}, s_{\tau}, s_{<\tau})$, then an expected (predictable) event has occurred. 

VPR has been shown to exhibit the desired properties pertaining to a hierarchical cognitive model -- the emergence of nested subjective timescales and segmentation of sequential signal into semantically coherent and hierarchically structured sub-sequences. Specifically, VPR is able to represent progressively `slower' features in the higher levels of its latent hierarchy, thus preserving the temporal organisation of the dataset's factors of variation. As such, this work explores an interesting avenue of research that is intriguingly connected to the field of time perception. In particular, the unexpected event criterion can be regarded as a measure of visual salience of sensory signal which has been linked to human duration judgments \citep{Sherman:2020, Fountas:2021}, while both of the criteria (in their function of detecting anticipated and unanticipated events) seem to draw a connection with the neuromodulatory attention mechanisms in the brain \citep{Yu:2005}.

\bibliographystyle{agsm}
\bibliography{library}
\end{document}